\documentclass{emulateapj}

\begin{document}

\shortauthors{Gordon et al.}
\shorttitle{Star Formation in M81}

\title{Spatially Resolved Ultraviolet, H$\alpha$, Infrared, and Radio
Star Formation in M81}  

\author{K.~D.~Gordon\altaffilmark{1}, 
   P.~G.~P\'erez-Gonz\'alez\altaffilmark{1}, 
   K.~A.~Misselt\altaffilmark{1}, 
   E.~J.~Murphy\altaffilmark{2},
   G.~J.~Bendo\altaffilmark{1}, 
   F.~Walter\altaffilmark{3},
   M.~D.~Thornley\altaffilmark{4,5},
   R.~C.~Kennicutt, Jr.\altaffilmark{1},
   G.~H.~Rieke\altaffilmark{1}, 
   C.~W.~Engelbracht\altaffilmark{1}, 
   J.-D.~T.~Smith\altaffilmark{1}, 
   A.~Alonso-Herrero\altaffilmark{1}, 
   P.~N.~Appleton\altaffilmark{8},
   D.~Calzetti\altaffilmark{5},
   D.~A.~Dale\altaffilmark{6},
   B.~T.~Draine\altaffilmark{7},
   D.~T.~Frayer\altaffilmark{8},
   G.~Helou\altaffilmark{8},
   J.~L.~Hinz\altaffilmark{1}, 
   D.~C.~Hines\altaffilmark{1,9},
   D.~M.~Kelly\altaffilmark{1}, 
   J.~E.~Morrison\altaffilmark{1},
   J.~Muzerolle\altaffilmark{1},
   M.~W.~Regan\altaffilmark{5},
   J.~A.~Stansberry\altaffilmark{1}, 
   S.~R.~Stolovy\altaffilmark{8},
   L.~J.~Storrie-Lombardi\altaffilmark{8},
   K.~Y.~L.~Su\altaffilmark{1}, and
   E.~T.~Young\altaffilmark{1}
   }
\altaffiltext{1}{Steward Observatory, University of Arizona, Tucson, AZ 85721}
\altaffiltext{2}{Dept.\ of Astronomy, Yale University, New Haven, CT 06520}
\altaffiltext{3}{NRAO, PO Box O, Socorro, NM 87801}
\altaffiltext{4}{Dept.\ of Physics, Bucknell University, Lewisburg, PA 17837}
\altaffiltext{5}{Space Telescope Science Institute, Baltimore, MD 21218}
\altaffiltext{6}{Dept.\ of Physics \& Astronomy, Univ.\ of Wyoming,
   Laramie, WY 82071}
\altaffiltext{7}{Princeton University Observatory, Princeton, NJ 08544}
\altaffiltext{8}{Spitzer Science Center, Caltech, Pasadena, CA 91125}
\altaffiltext{9}{Space Science Institute, Boulder, CO 80301}

\begin{abstract} 
We present Multiband Imaging Photometer for {\em Spitzer} (MIPS)
observations of M81 at 24, 70, and 160~$\micron$.  The grand design
nature of M81 is clearly seen, showing two well resolved
spiral arms containing numerous bright star forming regions.  The MIPS
images reveal a significant amount of cold dust associated with the
spiral arms.  We investigate the variation of the ultraviolet (UV),
H$\alpha$, and infrared (IR) luminosities and star formation rate (SFR)
indicators across the face of M81 using the MIPS images and archival
UV and H$\alpha$ images.  For regions in M81, we find that UV and
H$\alpha$ SFRs (uncorrected for dust attenuation) are always lower
than the IR SFR.  The cause of this behavior is dust attenuation
and/or using SFR calibrations appropriate for entire galaxies, not
regions in galaxies.  The characteristics of the dust attenuation for
the regions indicate the dust grains and/or geometry are different
from those in starburst galaxies.  The behavior of the infrared-radio
correlation in M81 is seen to vary from the global average, with
variations correlated with the morphology of M81.
\end{abstract}


\section{Introduction}
\label{sec_intro}

The galaxy M81 (NGC 3031) is one of the largest grand design spiral
galaxies in the sky ($14\arcmin \times 27\arcmin$).  It is a member of
the M81 galaxy group whose principle members are M81, M82, NGC 3077,
and NGC 2976 and include a number of dwarf galaxies \citep{Appleton81,
Boyce01}.  Prominent tidal HI tails around M81 show that it is
interacting with M82 and NGC3077 \citep{Yun94}.  The Cepheid distance
to M81 is 3.6~Mpc \citep{Freedman01}.  The proximity of M81 has made
it a favorite target for many investigations of galaxy properties from
the X-ray to radio \citep[e.g.][]{Kaufman87, Devereux95, Hill95,
Kong00, Swartz03}.

Many, well-resolved images of M81 from ultraviolet (UV) to radio
wavelengths have been taken, with the notable exception of the
infrared (IR).  Previously, well resolved far-IR images of galaxies
have only been possible for Local Group galaxies such as the Magellanic
Clouds \citep{Braun98, Wilke03}, M31 \citep{Haas98}, and M33
\citep{Hippelein03}.  With the successful launch of the {\em Spitzer
Space Telescope} \citep{Werner04}, it is now possible to map many
large galaxies in the far-IR with good spatial resolution, good
sensitivity, and in a reasonable amount of time.   Infrared Array
Camera \cite[IRAC,][]{Fazio04} and the Multiband Imaging Photometer
for {\em Spitzer} \citep[MIPS,][]{Rieke04} images of M81 were taken as part
of the commissioning of {\em Spitzer} and showcased in the first {\em
Spitzer} press release.  The MIPS images are presented in this paper
and \citet{Willner04} presents the IRAC images.  M81 is one of the key
galaxies in the {\em Spitzer} Infrared Nearby Galaxies
Survey\citep[SINGS,][]{Kennicutt03}.

We have chosen to investigate the variation of star formation rate
(SFR) indicators and the infrared-radio correlation across the disk of
M81 to highlight two of the questions which can be probed with these
new observations.  We compare the MIPS observations with existing UV
and H$\alpha$ images to probe the behaviors of the IR, H$\alpha$, and
UV SFR indicators across M81.  Such comparisons of SFR indicators have
been made for global galaxy fluxes
\citep[eg.,][]{Bell01, Kewley02, Rosa02}, but rarely has it been
possible to resolve all three SFR indicators in a single galaxy.  By
studying the resolved behavior of these SFR indicators, a better
understanding of their sensitivities to dust and stellar age can be
determined.  This will improve the accuracy of these SFR indicators
for both resolved and global galaxy measurements.  Radio emission is
another possible SFR indicator, but it is not as mature as the others
\citep{Haarsma00, Bell03}.  As such, we have chosen to
explore the more established infrared-radio correlation
\citep{Helou85, Condon92, Yun01, Pierini03}.

\section{Data}

\begin{figure*}[tbp]
\includegraphics[scale=1.1,angle=90]{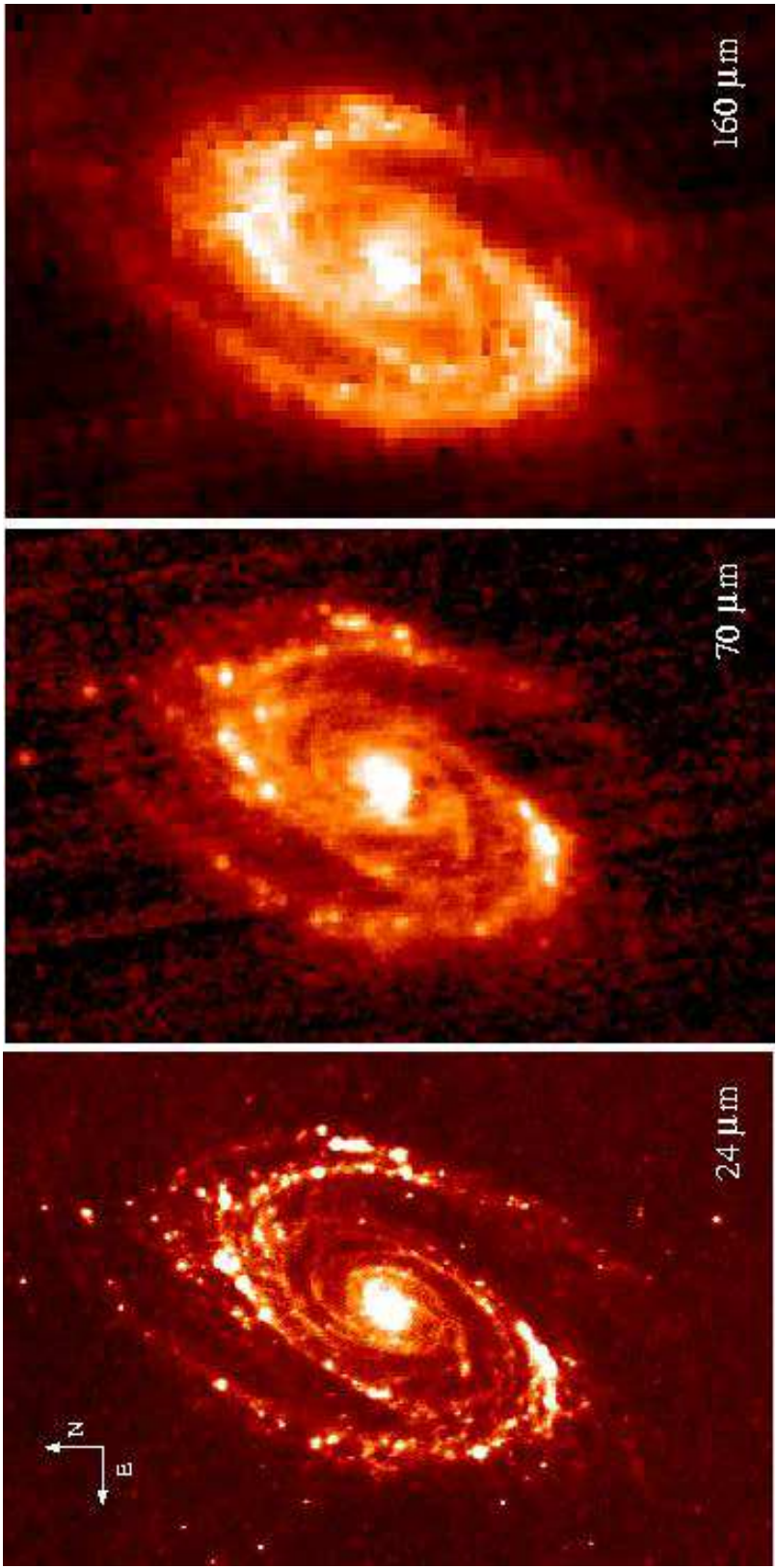}
\caption{The MIPS images of M81 are shown at full instrumental
resolution.  These resolutions are 6$\arcsec$, 18$\arcsec$, and
40$\arcsec$ for 24, 70, and 160~$\micron$, respectively.  The
field-of-view of the images is $20\arcmin \times 30\arcmin$ which is
more than large enough to cover M81.  These  
mosaics have a total exposure time of approximately 80, 40, and
8~seconds per point at 24, 70, and 160~$\micron$, respectively.
\label{fig_mips_fullres} }
\end{figure*}

\begin{figure*}[tbp]
\includegraphics[scale=1.1,angle=90]{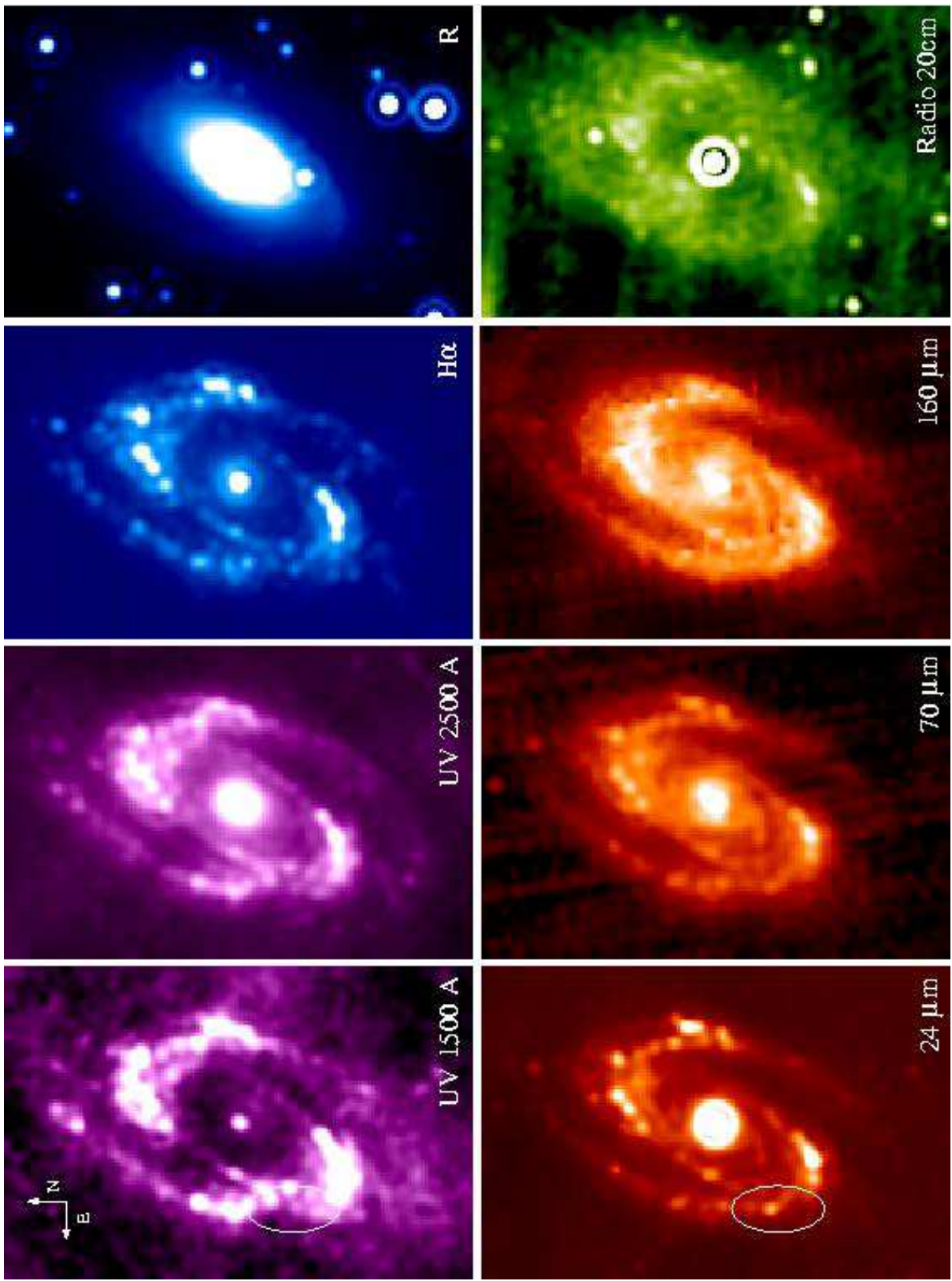}
\caption{The MIPS and ancillary images of M81 are shown convolved 
to have the 160~$\micron$ PSF which introduces a prominent Airy ring
around bright point sources.  The field-of-view of the images is
$15\arcmin \times 23\arcmin$.  Bright, foreground stars were masked in
the UV and H$\alpha$ images prior to convolution.  The ellipse denotes
the region discussed in \S\ref{sec_morph}.
\label{fig_all_images} }
\end{figure*}

\subsection{MIPS Observations}
\label{sec_mips_obs}

Images of M81 at 24, 70, and 160~$\micron$
were obtained in a
manner similar to that described in \citet{Engelbracht04} for NGC~55.
The MIPS images were reduced using the MIPS Instrument Team Data
Analysis Tool \citep{Gordon04} as described in
\citet{Engelbracht04}.  The uncertainties on the final absolute
calibrations are estimated at 10\%, 20\%, and 20\% for the 24, 70, and
160~$\micron$ data, respectively.  The final MIPS mosaics are
displayed in Fig.~\ref{fig_mips_fullres}.  The 70 and 160~$\micron$
images exhibit linear streaks along the scan direction (roughly
north-south) which are residual instrumental artifacts due to the time
dependent responsivity of the Ge detectors.
 
\subsection{Ancillary Data}

Ultraviolet images of M81 were taken with the Ultraviolet Imaging
Telescope in far-UV and near-UV bands \citep{Stecher97, Marcum01}.
The deepest, geometrically corrected, flux calibrated images
for each band were retrieved from the Multimission Archive at Space
Telescope.  The H$\alpha$ image of M81 was taken from
\citet{Greenawalt98}.  The radio image at 20~cm was created from
archival Very Large Array (VLA) C--array data, using standard
AIPS reduction techniques.

\subsection{Resolution Matching}

Each image of M81 was cropped to a common field-of-view ($20\arcmin
\times 30\arcmin$) and convolved to match the 
point-spread-function (PSF) of the MIPS 160~$\micron$ image to allow
for a consistent comparison of the images.  We used a Fourier
Transform technique for the convolution as described in
\citet{Engelbracht04}.  All of the convolved images
are shown in Fig.~\ref{fig_all_images}.  The 160~$\micron$ PSF clearly
has an Airy ring resulting in a prominent ring surrounding point
sources in some of the images.

\section{Results}

\subsection{Morphology}
\label{sec_morph}

At 160~$\micron$ resolution, the grand design nature of M81 is very
prominent in all but the R band image.  The R band image displays a
morphology dominated by the bulge, with only faint spiral arms
present.  At the other wavelengths, the spiral arms show a continuum
of morphologies from being a string of point sources (H$\alpha$ and
24~$\micron$) to exhibiting both point sources and an underlying
smoother distribution (UV and 70~$\micron$) to fairly smooth arms with
faint point sources (160~$\micron$ and radio).  This difference in the
appearance is related to the age of stars probed and the importance of
dust attenuation.  For example, H$\alpha$ probes younger star
formation than the UV images and, as a result, the UV images are
smoother than the H$\alpha$ image.  Evidence that dust attenuation is
affecting the arm morphology is seen in the southeast of M81 where a
prominent piece of the galaxy is missing from the UV and H$\alpha$
images (see ellipse in Fig.~\ref{fig_all_images}), but not from the IR
or radio images.

\subsection{Star Formation Rate Indicators}

\begin{figure}[tbp]
\epsscale{1.0}
\plotone{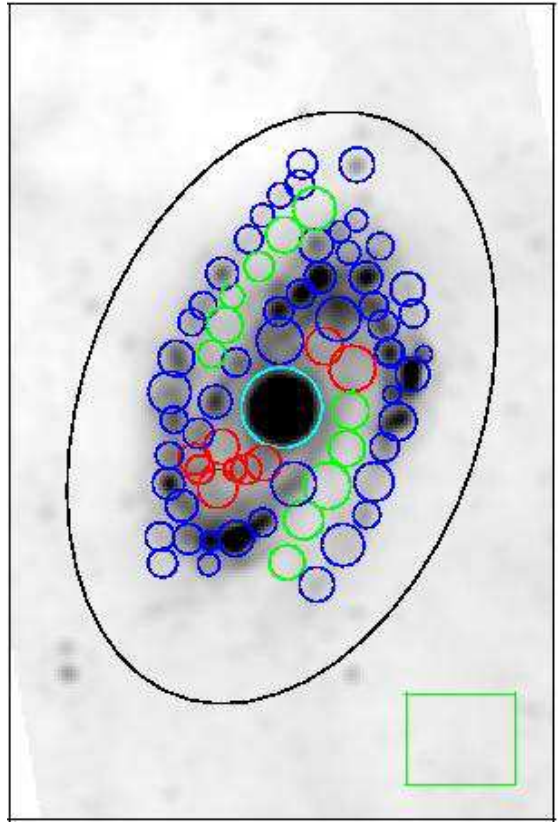}
\caption{The regions for which fluxes were extracted are shown
superimposed on the M81 24~$\micron$ image.  The size of the
extraction aperture is reflected in the size of the circle for each
region.  The average diameter of the apertures is 1.2~kpc.  The color
coding of the regions is given in Fig.~\ref{fig_sfr_reg}.  The green
box in the lower right corner gives the sky aperture.  At
70~$\micron$, the sky noise contribution to the uncertainty in the
average aperture is $\sim$2~mJy.  This shows that the linear streaks
are not affecting the 70~$\micron$ aperture measurements as all the
fluxes are above 80~mJy.  The field-of-view of the image is $20\arcmin
\times 30\arcmin$.
\label{fig_ext_reg} }
\epsscale{1.0}
\end{figure}

We investigated the behavior of the popular UV, H$\alpha$, and
IR SFRs in M81.  H$\alpha$ and UV SFR indicators suffer from dust
attenuation and probe stellar ages of 10 and 100~Myr, respectively.
The IR SFR indicator does not have an age bias, but does
assume that all of the starlight from the star formation is
absorbed by dust and re-emitted in the IR.  As we use resolved
images of M81 for all three SFR indicators, we can probe the biases
affecting each method in detail.  We have extracted fluxes for specific
regions in M81 from the 160~$\micron$ resolution images, correcting
the UV and H$\alpha$ fluxes for a foreground Milky Way extinction with
an $E(B-V) = 0.08$ \citep{Schlegel98}. The regions are shown in
Fig.~\ref{fig_ext_reg}.  

\begin{figure*}[tbp]
\plottwo{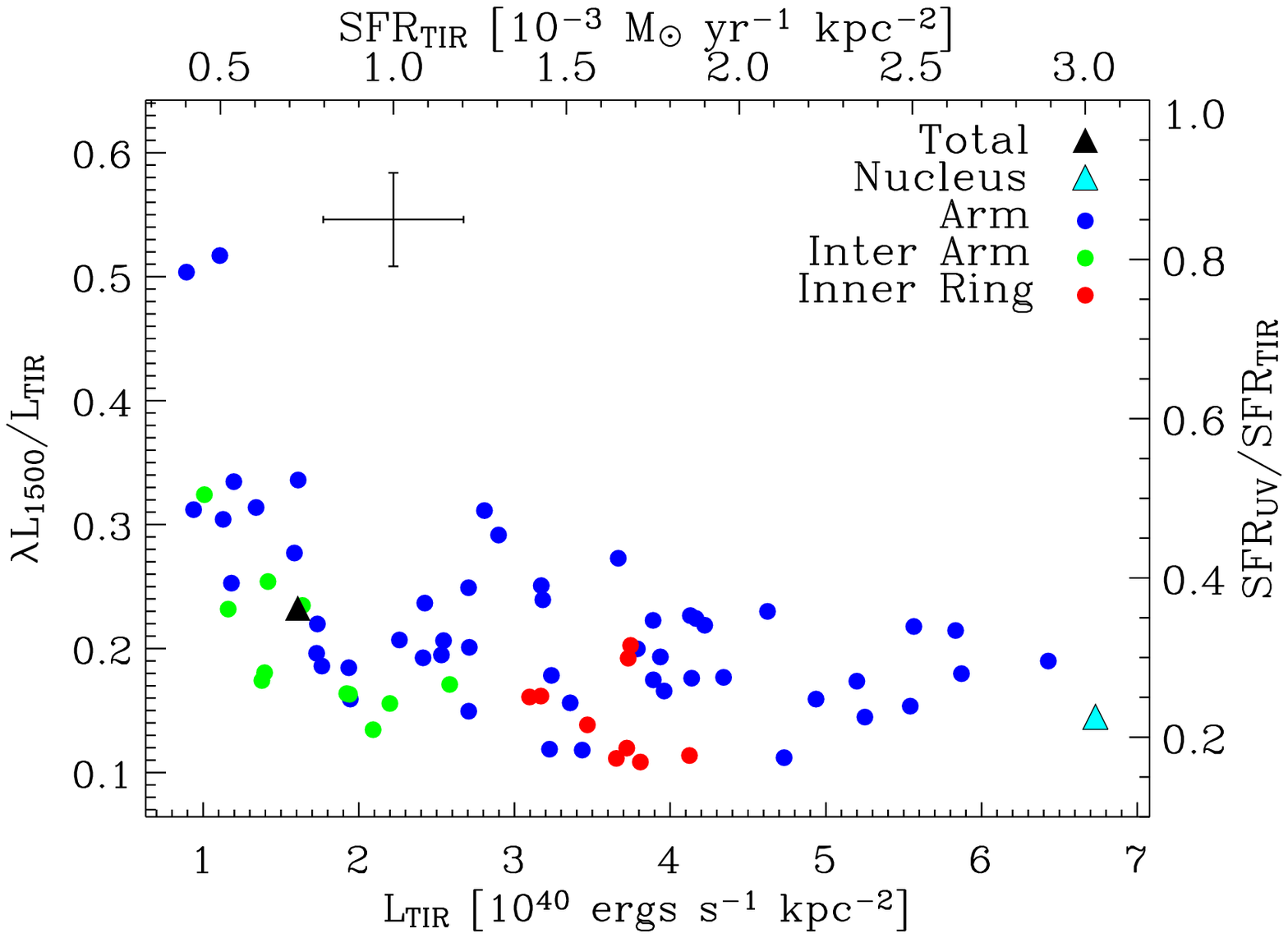}{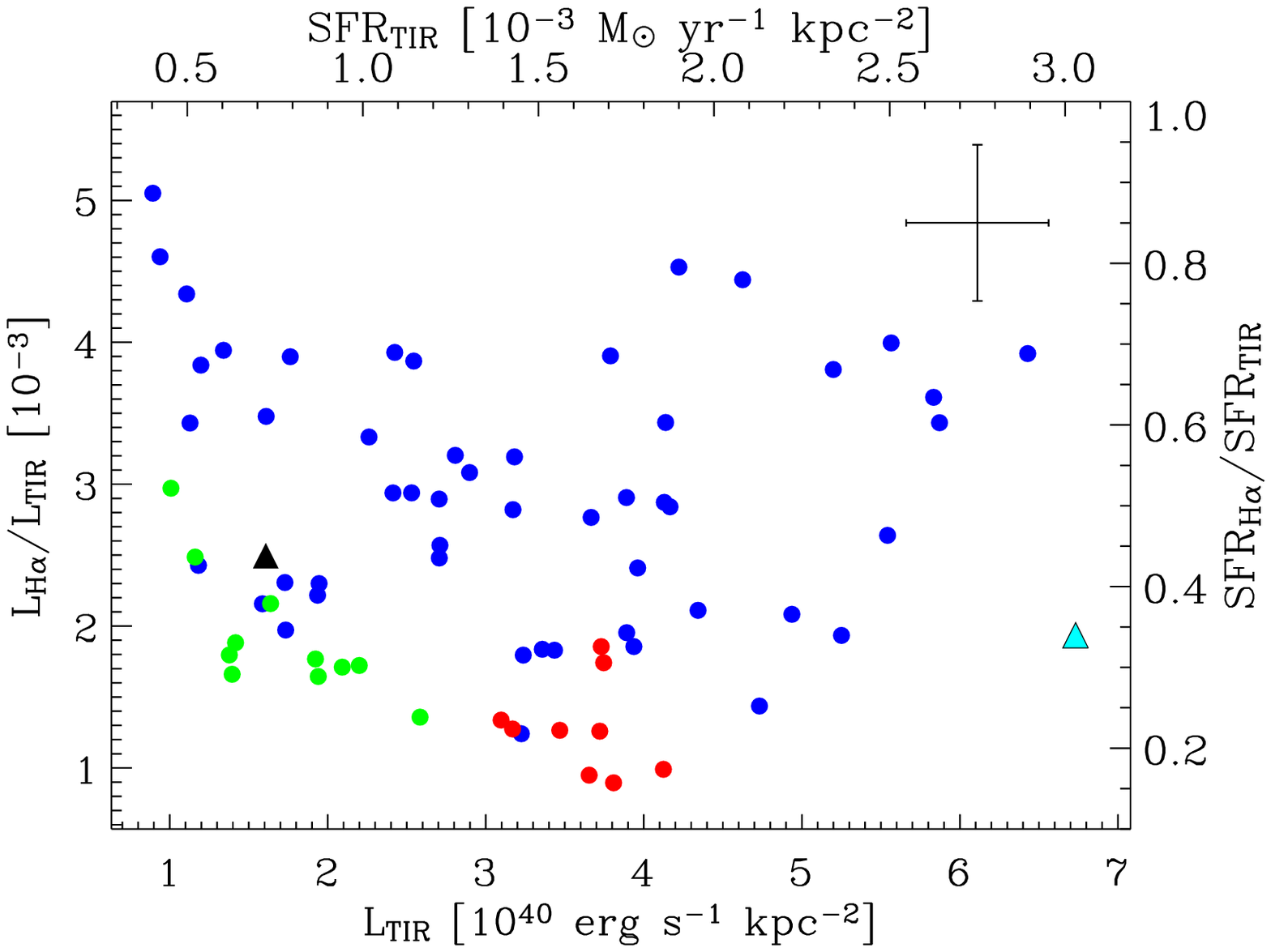}
\caption{The luminosities for UV and H$\alpha$, ratioed to the
IR luminosity, are plotted versus the IR luminosity.  The
IR luminosity is normalized to the area in $kpc^2$ of
each extraction aperture.  The corresponding SFRs and SFR ratios are
indicated on the opposite axes.  Representative uncertainties are shown.
\label{fig_sfr_reg} }
\end{figure*}

In order to convert measured UV, H$\alpha$, and IR luminosities
to SFRs, calibrations are needed.  One way to determine such
calibrations is to use stellar evolutionary synthesis models
\citep[eg.,][]{Fioc97,Leitherer99}.  \citet{Kennicutt98} has provided 
such calibrations with stellar age assumptions appropriate for entire
galaxies.  While the \citet{Kennicutt98} calibrations will be invalid
at some level for regions of galaxies, we have chosen to use them as
they allow a physical basis for interpretation of luminosities.  The
total IR flux was calculated using the TIR equation of \citet{Dale02}
formulated specifically to determine the 3--1100~$\micron$ luminosity given
MIPS measurements.  All of the SFRs in this paper are uncorrected for
dust attenuation as proper treatment of dust attenuation is beyond the
scope of this paper \citep[eg.,][]{Witt00}.

The SFRs for the whole galaxy are estimated to be 0.31, 0.38, and 0.89
$M_{\sun}/yr$ from the integrated UV, H$\alpha$, and IR
luminosities, respectively.  The 
luminosities and SFRs determined in the regions defined in
Fig.~\ref{fig_ext_reg} are plotted in Fig.~\ref{fig_sfr_reg}.  The arm
regions display the largest area-normalized IR SFRs
which is evidence that they have the highest SFRs.  The ratio of
H$\alpha$/IR luminosities is higher for the arm regions than any of
the other regions.  For all the regions, the UV and H$\alpha$ SFRs are
always smaller than the IR SFRs and there is a weak trend
towards smaller UV/TIR and H$\alpha$/TIR SFR ratios as the TIR SFR
increases.  There 
are three possible causes of these effects: 1) Dust is attenuating the
UV and $H\alpha$ measurements.  2) A fraction of the IR flux may not
be associated with recent star formation (i.e., dust heated by old
stars).  3) The luminosity-SFR calibrations of \citet{Kennicutt98} are
not appropriate for individual regions in a galaxy, where there may be
transfer of energy between different regions (radiative transfer
effects) and/or stellar age differences from those assumed by
\citet{Kennicutt98} which will affect SFR indicators at different
levels.  We will investigate the relative importance of these factors
in future papers but briefly discuss the first two points below.

\begin{figure}[tbp]
\plotone{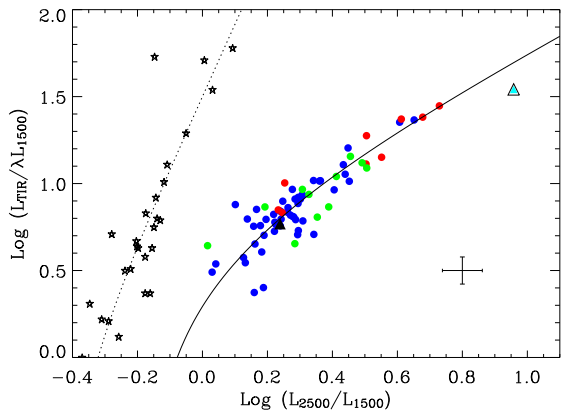}
\caption{The IR/1500 luminosity ratio is plotted versus the UV color
($L_{2500}/L_{1500}$ ratio).  The symbols are the same as in
Fig.~\ref{fig_sfr_reg}.  The solid curve gives the linear fit
[$L_{TIR}/\lambda L_{1500} = -2.42 + 3.81 (L_{2500}/L_{1500})$] to the
M81 data.  The star symbols give the observations for a sample of
starburst galaxies and the dotted line gives the \citet{Meurer99}
relationship determined for starburst galaxies.  Representative
uncertainties are shown.
\label{fig_dust_atten} }
\end{figure}

The association of significant IR emission with bright UV
and H$\alpha$ regions is evidence for significant dust attenuation.
The dust attenuation is probed in Fig.~\ref{fig_dust_atten} where the
IR/UV ratio and UV color attenuation measures are plotted.  The IR/UV
ratio has been shown to be a good dust attenuation measure for
galaxies and the UV attenuations for the extracted regions vary
between 0.5 and 3.0 mag for log($L_{TIR}/\lambda L_{1500}$) between
0.5 and 1.5 \citep{Gordon00}.  The $L_{2500}/L_{1500}$ ratio is also a
dust attenuation measure, but as it is a UV color, it is significantly
more uncertain \citep{Witt00}.  This flux ratio is very similar to the
more familiar $\beta$ dust attenuation measure \citep{Meurer99}.  The
starburst calibration of these measures is shown in
Fig.~\ref{fig_dust_atten} and does not describe the measurements of
M81.  Deviations from the starburst calibration
were also found by \citet{Bell02} for Large Magellanic Cloud HII
regions and by \citet{Kong04} for global measurements of galaxies more
quiescent than starbursts.  This implies that the dust geometry and/or
dust grain properties are different between starburst galaxies and
star forming regions in normal galaxies.

In studies of IR SFRs, it is common to assume that the cold dust is
mainly heated by old stars, and thus the cold dust emission
(160~$\micron$) should not be used in determining IR SFRs.  The M81
images presented in this paper allow us to probe this assumption for
the M81 disk but not the AGN dominated nucleus.  The morphology of
the spiral arms progresses from point-like to fairly smooth from 24 to
160~$\micron$.  The H$\alpha$ and UV bright regions are seen at
160~$\micron$ implying that young stars heat a fraction of the
160~$\micron$ emission.  The morphology of the IR images best matches
the H$\alpha$ and UV images as opposed to the R band image (an old
star tracer in M81), but this correlation is not enough to identify
the dominant 160~$\micron$ heating source, as the spiral density wave
concentrates the dust in the arms.  The smoother morphology of the UV
images compared to the H$\alpha$ image shows that the non-ionizing UV
bright stars have migrated out of their natal regions.  Young, hot
stars are more efficient than the old stars at heating dust since
their emission peaks at UV/blue wavelengths where dust is most
efficient at absorbing photons.  Given that both the cold dust and the
stars most efficient at heating dust are concentrated in the spiral
arms, it is likely that those stars are the dominant heating source
for the 160~$\micron$ dust in the spiral arms.

\subsection{Infrared-Radio Correlation}

\begin{figure}
\plotone{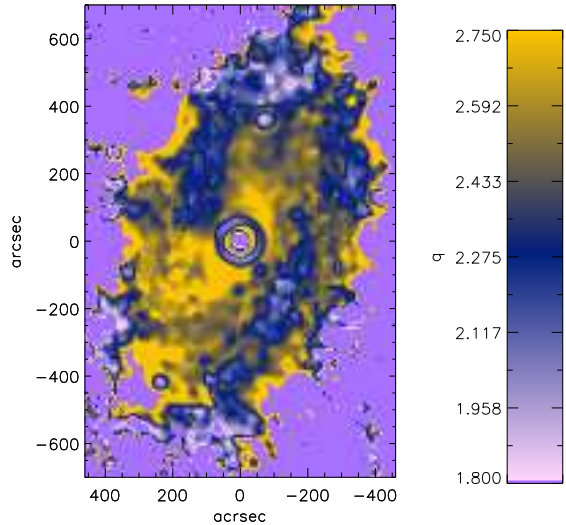}
\caption{The FIR to radio ratio ($q$) map is shown.  This map probes
the galaxy disk with a physical scale of $\sim 645$pc.  The field-of-view of
the image is $15\arcmin \times 23\arcmin$.  
\label{fig_radio_ir} }
\end{figure}

The far-IR dust emission and the radio continuum emission both trace
active star formation in galaxies, and are known to be tightly
correlated on galactic scales \citep{deJong85,Helou85}.  In
Fig.~\ref{fig_radio_ir}, we show the far-IR (42 - 122$\mu$m, FIR) to
radio ratio map of M81 constructed using the 160~$\micron$ resolution
MIPS images and 20~cm continuum data ($q \equiv {\rm log}({\rm
FIR}/3.75\times10^{-12} {\rm Hz}) - {\rm log}(S_{1.49{\rm GHz}})$).
The MIPS bands were combined to create a TIR image and the fractional
FIR \citep{Helou85} component was estimated per pixel using the
\citet{Dale02} models.  The value of $q$ ranges from $\sim$1.80 for
the AGN dominated nuclear region to $\sim$2.87 for the bright star
forming regions, with an intermediate value of $\sim$2.07 for the
interarm regions.  A similar range in $q$ was seen for M33 by
\citet{Hippelein03}, although they did not comment on any large scale
structure seen in $q$ variations as we see in M81.  For comparison,
the average $q$ value found for galaxies by \citet{Yun01} is 2.34.
The variation in $q$ is coherent with structures related to the M81
spiral arms, confirming that the radio-infrared correlation is more
complex inside individual galaxies than between galaxies.  This
variation will be investigated in detail in future papers to explore
the dynamics of cosmic-ray electron propagation and escape
\citep{Helou93}.

\section{Conclusions}

We presented MIPS images of M81 at 24, 70, and 160~$\micron$ which
reveal a bright nucleus and two well resolved spiral arms studded with
bright regions of star formation.  These images show that M81 has a
significant amount of cold dust associated with the spiral arms.  From
multiwavelength morphology comparisons, the dust heating is argued be
dominated by recent star formation even at the longest MIPS
wavelength.  The resolved UV and H$\alpha$ SFRs are always lower than
the IR SFRs indicating significant dust attenuation, radiative
transfer effects, and/or different stellar ages than assumed for the
\citet{Kennicutt98} SFR calibrations.  The character of the dust
attenuation indicates that the dust geometry and/or grain properties
are different for resolved regions in M81 than in starburst galaxies.  The
infrared-radio correlation was found to vary by a factor of $\sim$6 in
the M81 with coherent structures relating to the spiral arms.  These
results illustrates the need for additional theoretical and empirical
work on how to accurately combine different SFR indicators, accounting
for dust, age, and radiative transfer effects, to give an accurate
view of the star formation in regions of galaxies.

\acknowledgements
The SINGS team is warmly thanked for allowing M81 to be part of
the ERO program.  We thank Rene Waterbos for providing us with the
H$\alpha$ and R band images.  The radio image used in this paper was
taken with the VLA which is operated by the National Radio Astronomy
Observatory.  This work is based on observations made
with the {\em Spitzer Space Telescope}, which is operated by the Jet
Propulsion Laboratory, California Institute of Technology under NASA
contract 1407. Support for this work was provided by NASA through
Contract Number \#960785 issued by JPL/Caltech.

\end{document}